\documentclass[eqsecnum,aps,prd,nofootinbib,twocolumn]{revtex4}
\usepackage{amsmath}
\usepackage{enumerate}

\begin{document}

\title {Gyrating Strings: A new instability of black strings?}
\author{Donald Marolf}

\affiliation{Physics Department, UCSB, Santa Barbara, CA 93106, USA\\
{\tt marolf@physics.syr.edu}}

\author{Belkis Cabrera Palmer}
\affiliation{Physics Department, Syracuse University, Syracuse, NY 13244,\\
Physics Department, UCSB, Santa Barbara, CA 93106. \\
{\tt bcabrera@physics.ucsb.edu}}

\date{April, 2003}

\begin{abstract}
A thermodynamic argument is presented suggesting that
near-extremal spinning D1-D5-P black strings become unstable when
their angular momentum exceeds $J_{crit} = {3Q_1Q_5}/2\sqrt{2}$.
In contrast, the dimensionally reduced black holes are
thermodynamically stable. The proposed instability involves a
phase in which the spin angular momentum above $J_{crit}$ is
transferred to gyration of the string in space; i.e., to orbital
angular momentum of parts of the string about the mean location in
space.  Thus the string becomes a rotating helical coil.
We note that an instability of this form would yield a
counter-example to the Gubser-Mitra conjecture, which proposes a
particular link between dynamic black string instabilities and the
thermodynamics of black strings. There may also be other instabilities associated with radiation modes of various fields.   Our arguments also apply to the
D-brane bound states associated with these black strings in weakly
coupled string theory.
\end{abstract}

\maketitle

\section{Introduction}

Black holes remain one of the most intriguing objects in general
relativity, and are known for their simplicity and stability.
Above 3+1 dimensions, gravity can produce not only black holes,
but also extended black objects such as strings and p-branes.  In
many ways, these extended black objects behave like their  black
hole cousins.  Indeed, the most familiar examples of black strings
and branes have translational symmetries, and dimensional
reduction along these symmetries yields black hole solutions to
lower dimensional theories.

However, properties of black branes can sometimes differ
significantly from those of the associated black holes.  The {Gregory-Laflamme
instability \cite{GL,GLcharge,GLextreme} is a classic example of
such behavior.  In \cite{GL} it was shown that certain black
strings are dynamically unstable to a breaking of translational
(but not rotational) symmetry along the string.  The ultimate fate
of this instability remains a matter of investigation, but
conjectures \cite{GHKM} as to this fate have led to the discovery
of inhomogeneous black strings\cite{wiseman} (though these solutions
cannot be the endpoint of the Gregory-Laflamme instability, at
least in $d \le 13$ dimensions \cite{ES}; see also \cite{Kol})
 and related work on dynamics \cite{choptuik}.

The general theory of such instabilities remains to be understood.
An oft-discussed conjecture in this context was stated by Gubser
and Mitra \cite{GM}, who proposed that black branes might have
dynamical instabilities precisely when they have thermodynamic
instabilities, in the sense that the Hessian of second derivatives
of the energy with respect to the conserved charges has negative
eigenvalues.  This conjecture has been proven in certain contexts \cite{Harvey}.

Here we argue for an instability which places a new twist on such
discussions.  We examine the three-charge (D1-D5-P) spinning BPS
black brane of type IIB supergravity, which becomes a 5+1 black
string when compactified on $T^4$.  When compactified along the
remaining translation symmetry and dimensionally reduced to 4+ 1
dimensions, the resulting black hole is dual to that studied in
\cite{BMPV} by Breckenridge, Myers, Peet and Vafa  and is a
rotating version of the black hole whose entropy was counted
\cite{SV} by Strominger and Vafa using D-brane techniques.  As in
the works above, we take the direction along the string to be
compactified in a circle of length $L$ in order to yield finite
charges. The near-extremal solution was studied in
\cite{nearBMPV,CY1,CY2}.   In particular, from the results of
\cite{CY1} one can show that it has no thermodynamic instabilities in the
sense of Gubser and Mitra\footnote{In addition, the microscopic
entropy of such objects in string theory was computed in \cite{CL}
using AdS/CFT methods.}.   
Thus their conjecture predicts dynamical
stability\footnote{However, as will be discussed later, there is one direction which the system is thermodynamically only marginally stable.}.

However, this theory contains other BPS black strings carrying the
same charges. In particular, strings were described in \cite{HM2}
in which all or part of the angular momentum is carried by {\it
gyrations} of the string as opposed to spin.  In such solutions
the black string at any instant of time can be thought of as being
helical in space, with the helical profile traveling along the
string at the speed of light.  Such solutions are easily generated
by applying the technique of Garfinkle and Vachaspati \cite{GV} to
the spinning strings of \cite{BMPV}.   In fact, the shape of such
gyrations need not be a helix, but can be much more general.  As a
result, the space of such BPS black branes is highly degenerate;
for fixed charges, one must still specify several functions on a
circle in order to determine the solution uniquely.   Oscillating
versions of similar strings had previously been constructed in
\cite{CMP,DGHW,LW,CT}.

Below we study the general gyrating BMPV string with anti-self
dual angular momentum and find that, for large enough angular
momentum ($J
> J_{crit} =3 Q_1Q_5/2\sqrt{2}$), it is entropically favorable for
the excess angular momentum $J-J_{crit}$ to be carried by
gyrations. In particular, maximizing the entropy $S$ over the
class of solutions carrying anti-self dual spin angular momentum
$J_{spin}$ and anti-self dual gyrational angular momentum
$J_{gyro}$, we find that for $J > J_{crit}$ the entropy $S$ is an
increasing function of $J_{gyro}$ on the interval $[0,
J-J_{crit}]$, with positive slope for $0\le J_{gyro} < J_{crit}$.
This suggests that a small perturbation of a non-gyrating black
string could grow to become large; i.e., it suggests an
instability of the black string\footnote{The discussion in
\cite{HM2} explicitly focussed on the case where the gyrations are
small enough that the object could be well described via
dimensional reduction to a black {\it hole}. In such cases, it was
shown that the effect of gyrations on the black string entropy is
negligible. However, the gyrating string has significantly larger
entropy only when $J-J_{crit}$ is of order $J$, and in this case
the oscillations are necessarily large.}.

Some useful details of the solutions are described in section
\ref{details} below, as they were suppressed in \cite{HM2}.  Section \ref{entropy} then assembles the argument and
comments on the importance of small deviations from extremality.
We also discuss two possible scenarios for the final fate of this
instability.  Section \ref{disc} contrasts our situation with that associated with super-radiance, observes that an analogous argument holds for the D-brane bound states associated with our black strings in weakly coupled string theory, and makes further final comments.

\section{The gyrating black string}
\label{details}

Consider a D1-D5-P black brane solution which is asymptotically
$R^5 \times S^1 \times T^4$.  We will think of the $T^4$ as being
small so that the solution is effectively a black string in 5+1
dimensions.  For a translationally invariant brane, the
ten-dimensional type IIB supergravity solution is

\begin{eqnarray}
\label{straight} ds^2 &=& \left( 1+\frac{r_o^2}{r^2}
\right)^{-1}(-dudv+\frac{p}{r^2}du^2 \nonumber \\
&+& \frac{2\gamma}{r^2}(\sin^2 \theta d\varphi - \cos^2 \theta d
\psi) du) \nonumber \\ &+& \left( 1+\frac{r_o^2}{r^2} \right) dx_i
dx^i + dy_ady^a.
\end{eqnarray}
The index $a$ runs over the four directions of the $T^4$ and $i$
runs over the four space directions transverse to the branes.
These latter 4 directions are associated with an $S^3$ labeled by
constant values of $r^2 = \sum_i x^ix^i$.  The angles $\theta,
\varphi, \psi$ label this $S^3$, while the coordinates $z,t$ label
the worldvolume directions of the string. Here we have chosen the
special case where D1- and D5-brane charges are tuned to achieve
constant dilaton and four-torus volume, so that (\ref{straight})
is the metric in either the string or Einstein frame. In
particular, the D1- and D5-charges are

\begin{equation}
Q_1 = \frac{V r_0^2}{(2\pi)^4 g}, \ \ Q_5 = \frac{r_0^2}{g},
\end{equation}
where $g$ is the string coupling and $V$ is the volume of the four-torus.
This solution has a null Killing vector field $\partial /
\partial v$, so one may attempt to add travelling waves via the method of \cite{GV}.
It turns out that there are many interesting such waves for this
system, which were studied extensively in \cite{LW,HM1,HM2}
following similar work of \cite{CMP,DGHW,CT} for other systems.

We are interested here in the class of waves briefly discussed in
\cite{HM2} which can be viewed as {\it gyrations} of the brane
itself in the $x^i$ directions.  Such solutions differ from
(\ref{straight}) only by the addition of a term proportional to $
\ddot{h}_i(u)x^i du^2$, where $h_i$ are arbitrary functions of
$u$:

\begin{eqnarray}
ds^2 &= & \left(
1+\frac{r_o^2}{r^2}\right)^{-1}(-dudv+\frac{p}{r^2}du^2 \nonumber
\\ &+& \frac{2\gamma}{r^4}( x^1 dx^2 - x^2 dx^1 - x^3dx^4 +
x^4dx^3) du-2\ddot{h}_i(u)x^idu^2)
\nonumber \\
&+& \left( 1+\frac{r_o^2}{r^2} \right) dx_i dx^i+dy_idy^i.
 \end{eqnarray}
After the change of coordinates

\begin{eqnarray}
v' &=&v+2 \dot{h}_i x^i + \int^u \dot{h}^2    du ,\\
x'^i &=& x^i+ h^i,
\end{eqnarray}

asymptotic flatness of the metric becomes manifest:

\begin{eqnarray}
\label{gyrate}
 ds^2 &=& \left(
1+\frac{r_o^2}{|\vec{x}-\vec{h}|^2}\right)^{-1}
\Bigl[-dudv+\frac{p}{|\vec{x}-\vec{h}|^2}du^2 \nonumber \\
&+& \frac{2\gamma}{|\vec{x}-\vec{h}|^4}(\Delta^1 d\Delta^2 -
\Delta^2 d\Delta^1 - \Delta^3d\Delta^4 + \Delta^4d\Delta^3 ) du
 \nonumber \\
&+&  \left( 2+\frac{r_o^2}{|\vec{x}-\vec{h}|^2}\right)
\frac{r_o^2}{|\vec{x}-\vec{h}|^2 } (\dot{h}^2 du^2 -2 \dot{h}_i
dx^i du )\Bigr]
\nonumber \\
&+& \left( 1+\frac{r_o^2}{|\vec{x}-\vec{h}|^2} \right) dx_i dx^i +
dy_idy^i,
 \end{eqnarray}
where $\Delta^i = x^i - h^i$.
Note that the black string horizon is located at $x^i = h^i(u)$ in the new coordinates,
suggesting that the string is indeed oscillating in the $x^i$
directions. Following \cite{HM2}, we have in mind circular
oscillations of the type associated with a net angular momentum and we
refer to {\it gyrations} of the string in the $x^i$ directions.

The conserved linear and angular momenta of the gyrating black
string can be read off from the asymptotic form of (\ref{gyrate}).
 The results are as follows:
 \begin{eqnarray} \label{P}
P &=& \frac{Lp}{\kappa^2} + P_{gyro}, \ \  P_{gyro} = \frac{2 r_0^2}{\kappa^2} \int_0^L  \dot{h}^2du,     \\
J_\phi &\equiv& J_{12}= J_{12}^{gyro} + \frac{L\gamma }{\kappa^2},            \\
J_\psi &\equiv& J_{34}= J_{34}^{gyro} - \frac{L\gamma }{\kappa^2},            \\
\label{J} J_{ij}^{gyro} &=& \frac{2r_0^2}{\kappa^2} \int_0^L (h_i
\dot{h}_j-h_j \dot{h}_i) du,
\end{eqnarray}
Here $\kappa^2 = (2 \pi)^5 g^2/V$. Note that the expressions for
$P_{gyro}$ and $J^{ij}_{gyro}$ are identical to those of a
material string with tension $2r_0^2/\kappa^2$.

Now, for any string the gyrational angular momentum is bounded by
a linear function of gyrational momentum. This result may be
derived in several ways. For example, one might note that the
expressions for $P_{gyro}$ and $J_{gyro}$ have the same form as
those of \cite{MNT} for the angular momentum and charge of a
supertube. Applying their argument here yields $|J_{12}| \leq
\frac{P_{gyro} L}{2\pi}$. Another argument notes that quantizing
the string will yield vector particles, each carrying spin $\pm
1$.  Thus, the angular momentum must be bounded by the number
$N_{gyro} = P_{gyro}L/2\pi$ of associated momentum quanta.

However, requiring our angular momentum to be anti-self dual is
not compatible with saturating this bound.  As a result, the bound on our
angular momenta is even more stringent.  A
lengthy but otherwise straightforward application of variational
methods shows that, for fixed $N_{gyro}$, the largest anti-self
dual $J_{gyro}$ is obtained by placing $N_{gyro}/3$ excitations in
the fundamental ($k=1$) mode of the string and using them to carry
$J_{12} = N_{gyro}/3$, while also placing $N_{gyro}L/3$ excitations in
the $k=2$ mode and using it to carry $J_{34} = N_{gyro}/3$ and
$2N_{gyro}/3$ momentum quanta.  This configuration has $J_{gyro} =
\frac{\sqrt{2}}{3}N_{gyro}$ units of angular
momentum\footnote{There is in fact a several parameter space of
configurations carrying the same anti-self dual $J_{gyro}$, but
there are no configurations with higher such $J_{gyro}$.}.

\section{Entropy and Instability}
\label{entropy}

In \cite{HM2} it was shown that adding such gyrations does not
affect the induced metric on the horizon, and in particular that
the area of the horizon does not change\footnote{However, a mild
null shock-wave singularity does form along the horizon. The
metric is $C^0$ at the horizon (so that, in particular, its area
is well-defined) but it is not $C^1$.  One expects that any amount
of excitation above the BPS bound will smooth out this
singularity, though the resulting solution is unlikely to be
stationary.   See the appendix of \cite{HM1} for details of the
extreme solutions.}. Therefore, the entropy of the gyrating string
has the same form as that of the spinning D1-D5-P string
\cite{BMPV}, but with the total momentum replaced by $P -
P_{gyro}$ and the total angular momentum replaced by $J-
J_{gyro}$.  In terms of the number of momentum quanta $N=PL /
2\pi$ and $N_{gyro} = P_{gyro}L/2\pi$, the result is:
\begin{equation}
\label{entropy} S=2 \pi \sqrt{Q_1 Q_5 (N-N_{gyro})-(J-J_{gyro})^2}.
\end{equation}

Given a gyrating black string with $|J_{gyro}| < \frac{\sqrt{2}}{3} N_{gyro}$,
we can always decrease $N_{gyro}$ to
obtain a solution with larger entropy.  Thus maximally entropic
strings saturate this bound to yield:
\begin{equation}
\label{S2}
S=2 \pi \sqrt{Q_1 Q_5 (N-\frac{3}{\sqrt{2}}|J_{gyro}|)-(J-J_{gyro})^2} \\
\end{equation}

When the entropy is now maximized over $J_{gyro}$, the absolute
value in (\ref{S2}) leads to two distinct behaviors.  For $J <
J_{crit} := \frac{3Q_1Q_5}{2\sqrt{2}}$, the maximum is at $J_{gyro}
=0$.  However, for $J > J_{crit}$, entropy is maximized for
$J_{gyro} = J - J_{crit}$ and thus $J_{spin} = J_{crit}$.

We now use the above observations to argue for a new type of black
string instability.  Namely, we suggest that certain non-gyrating
spinning black strings are unstable to the development of gyration
for $J>J_{crit}$. Now, one does not expect BPS solutions to have a
linear instability\footnote{For static solutions, the BPS bound
and the results of \cite{SMB} work together to forbid linear
instabilities.  However, we know of no general theorems for the
stationary case.}.  Indeed, we have seen that they are marginally
stable to developing gyration, as any amount of gyration leads to
a stationary solution.  That is, the parameter $J_{gyro}$
effectively labels a moduli space of BPS solutions.

However, a generic perturbation of the spinning string will result in motion through this moduli space (as well as some amount of excitation of the string off of the moduli space).  The important observation is that near $J_{gyro}=0$ motion in the direction of increasing $J_{gyro}$ is entropically favored over motion in the direction of decreasing $J_{gyro}$.  In fact, we have seen that the entropy is minimized at $J_{gyro} =  J-J_{crit}$, so that this value is entropically stable.  As one may expect\footnote{Were the horizons completely smooth, this would be enforced by the area theorem \cite{Hawking,CDGH}.} this entropic stability to be enforced dynamically, we conjecture that the gyrating string with $J_{gyro}= J- J_{crit}$ is {\it dynamically} stable, perhaps due to higher order dynamical effects beyond the linear level.  Similarly, we conjecture that the spinning non-gyrating string with $J> J_{crit}$ is dynamically {\it unstable}, perhaps due to higher order effects.

Instead of considering BPS objects, one might consider strings
with energies slightly in excess of the BPS bound.  For a nearly
BPS object with $J$ substantially greater than $J_{crit}$, one
expects a similar entropy formula and a similar instability.   In
particular, if some form of continuity holds then we have that:

\begin{enumerate}
\item Near BPS solutions can also be labeled by a parameter
$J_{gyro}$.  Since one expects non-BPS gyrating strings to
radiate, these solutions are unlikely to be stationary.  Their
gyrating phase will be transient.  However, this means only that
$J_{gyro}$ will refer to the gyrational angular momentum at some
particular moment of time.

\item The derivative $\frac{\partial S}{\partial J_{gyro}}$ will
be positive at $J_{gyro}=0$, where the derivative is taken with
all conserved charges held fixed.
\end{enumerate}
Thus, one expects a near-BPS non-gyrating string with $J >
J_{crit}$ to be unstable to transfer of angular momentum from spin to
gyration.

It is interesting to speculate as to the final state into which
this string decays. There are two natural possibilities. The first
is that the string sheds its excess angular momentum through
classical radiation and eventually becomes a {\it stable}
non-gyrating string with $J \le J_{crit}$. The second is that the
dominant effect is shedding of excess {\it energy} and that the
final state is a gyrating {\it BPS} string. One might expect that
either final state can arise and that the outcome depends on the
particular values of the parameters. Note, however, that if we
were to place the unstable string in a small reflecting cavity,
this would prevent the loss of significant amounts of either $E$
or $J$, so that one would expect decay into a stable non-BPS
gyrating black string.  Due to the small domain and simple
boundary conditions, this might be a particularly interesting
arena for numerical simulations.  It would also lead to a clear
signal: an equilibrium state that is far from being rotationally
invariant.

\section{Discussion}

\label{disc}

We have argued for an instability of D1-D5-P near BPS black
strings with $J > J_{crit} = \frac{3Q_1Q_5}{2\sqrt{2}}$.  Note
that such strings exist only when the number $N$ of momentum
quanta exceeds a certain bound: $N \ge \frac{8}{9}Q_1Q_5$.   Thus,
like the original Gregory-Laflamme instability, the instability
arises only for sufficiently long strings.

Now, the thermodynamics noted above might also be taken to suggest an instability to simply radiating angular momentum (and momentum)
to infinity.  In fact, this latter sort of potential instability could in principle occur
at a smaller value of $J$, since gravitational waves can carry more $J$ for a given amount of $P$.
It is natural to take guidance from the study of 3+1 dimensional Kerr black holes, where one finds a similar thermodynamics: Kerr black holes have an entropy that decreases with increasing $J$.  Thus, radiation of a small amount of energy is allowed if it carries a large angular momentum.  In the case of Kerr one finds no instability for massless fields, but merely super-radiance: a given incident wave undergoes a finite amount of amplification and then disperses to infinity.  In contrast, a true linear instability does result \cite{PT,B1} if the black hole is surrounded by a large mirror (or is placed in a large AdS space \cite{B2}) so that the wave is continually redirected toward the black hole.

However, an instability for Kerr can arise for fields of massive fields, which can be bound to the black hole by the  gravitational potential. Results are known for minimally coupled scalar fields  \cite{DDR,SD,ZE,FN}.  In our context, one may expect radiation modes with momentum in the $z$-direction (i.e., Kaluza-Klein modes) to behave similarly, and our gyrational mode is much like such a bound state.  Indeed, in the non-BPS case it is not clear to us to what extent it can be meaningfully distinguished from such bound states.  But while a study of such bound states for non-BPS strings is difficult, it is clear the gyrational mode is the unique such bound state in the BPS limit.  Since this limit will be important below, we focus on the gyrational mode.

We have seen that gyrations cannot be re-absorbed into the black string since $\frac{\partial S}{\partial J_{gyro}} > 0$.
Thus, such gyrations can decay only through radiation to infinity.
Whether or not a linear instability occurs will then be determined by a competition between two effects: the amplification of the traveling wave and the tendency to radiate the gyrational traveling wave to infinity.  Both are expected to vanish in the BPS limit.  However, one expects the amplification to increase with $J-J_{crit}$, while there is no reason for this parameter to affect the rate at which the gyrational traveling wave is radiated to infinity.  Thus, one expects that, at least by tuning parameters so that $J-J_{crit}$ is large while the string remains nearly BPS, one can indeed produce an instability.

Our argument is highly suggestive, but clearly falls short of a proof.  We
have also described two possible final states.  The system clearly
calls for more detailed investigation and may yield a variety of
interesting phenomena. In addition to those mentioned above, it is
may also be fruitful to investigate relations between gyrating
black strings and black tubes or black rings \cite{ER,EE,Iosif}.

Supposing now that an instability (of any type discussed above) does occur. let us briefly reflect on the broader implications.  An interesting attempt to understand the general nature of black string instabilities is encoded in the Gubser-Mitra conjecture of \cite{GM}.
Quoting from \cite{GM}, this is the conjecture that

\begin{quote}
...for a black brane with translational symmetry, a Gregory-Laflamme instability exists precisely when the brane is thermodynamically unstable. Here, by ÒGregory-Laflamme instabilityÓ we mean a tachyonic mode in small perturbations of the horizon; and by Òthermodynamically unstableÓ we mean that the Hessian matrix of second derivatives of the mass with respect to the entropy and the conserved charges or angular momenta has a negative eigenvalue.
\end{quote}

This conjecture has been proven for a certain class of black
strings \cite{Harvey}, but our system appears to be a
counter-example to the conjecture holding in complete generality.
Let us consider a slightly non-BPS spinning string with $J >
J_{crit}$.  We choose the non-BPS case as, with asymptotically
flat boundary conditions, we expect gyrating strings to radiate so
that non-BPS non-gyrating strings will form an isolated family in
the space of stationary solutions.  Thus we may cleanly talk about
``the entropy of the black string with fixed conserved charges and
angular momenta.''  Nearly-BPS objects are generally
thermodynamically stable.  The details of the non-extremal solutions 
can be found in \cite{CY1}, and show that no instabilities are present near extremality.
Our
discussion above suggests a dynamical instability, and thus a counter-example to the above conjecture.
However, it should be noted that, even at a finite distance from extremality, one finds an interesting conspiracy that forces the Hessian to have a single zero eigenvalue.  Thus, it is possible that the conjecture may be preserved if non-extremal strings have a marginally stable mode leading to gyration, but no linear instabilities.

Shortly after the initial appearance of this work, another argument for a counter-example to Gubser-Mitra was given in \cite{RB} by exhibiting a marginally stable mode in a thermodynamically stable but near-BPS black string.  It is interesting to rephrase our results in the same terms:  on general grounds, one may expect that some Kaluza-Klein modes around rotating black strings are unstable as they should act like massive fields around Kerr black holes.  However, in general we expect such black strings to be thermodynamically unstable.  It is only in the BPS limit that one expects thermodynamic stability.  The gyrational mode of the BPS black string studied here allows us to see that a small number of marginal bound states persist for this string in the extremal limit, suggesting the presence of stable, marginal, and unstable modes within any neighborhood of the BPS limit, and in particular in the thermodynamically stable regime.

As a final comment, the reader may wish to return to the
discussion of \cite{BMPV} and ask how the entropy of gyrating BPS
black strings is to be understood from string theory.  The answer
is that gyration of a D-brane bound state is described by the U(1)
``center-of-mass'' degrees of freedom, since the D-branes gyrate
collectively in a way that does not excite the relative motion
degrees of freedom.  The counting of states for the spinning black
string given in \cite{BMPV} does not include the effect of this
degree of freedom as it is known to carry little entropy (see,
e.g., \cite{JuanThesis}).  Thus, momentum that goes into exciting
gyration of the bound state produces no entropy.  But this is just
what was observed in the black string entropy in equation
(\ref{entropy}).  The point is that the U(1) degrees of freedom
{\it can} carry angular momentum, and can do so more `cheaply'
than can the collective modes. Thus, allowing the U(1) degrees of
freedom carry linear momentum $P_{gyro}$ and angular momentum
$J_{gyro}$ leads directly to the entropy (\ref{entropy}) for the
gyrating D-brane bound state.  In particular, this means that for
certain values of the global charges D-brane bound states are also
unstable to gyration in the presence of interactions (e.g., via
closed-string exchange) between the center-of-mass U(1) and other
degrees of freedom.

\begin{acknowledgments}  This work was inspired by questions that arose from discussion with Iosif Bena.  We also thank Henriette Elvang, Eric Gimon,
Andr\'es Gomberoff, Gary Horowitz, Harvey Reall, and Rob Myers for other useful discussions.
This work was supported in part by NSF grants PHY00-98747,
PHY99-07949, and PHY03-42749, by funds from Syracuse University,
and by funds from the University of California. BCP was supported
in part by a Syracuse University Graduate Fellowship. In addition,
BCP thanks UCSB for its hospitality during this
work.

\end{acknowledgments}

\end{document}